\documentclass[final]{svjour3}

\usepackage[dvipdfmx]{graphicx}
\usepackage{rotating}
\usepackage{amssymb}
\usepackage{mathptmx}
\usepackage{siunitx}
\usepackage{xcolor}
\usepackage{physics}

\makeatletter
\journalname{Journal of Low Temperature Physics}
\graphicspath{ {/} }

\begin{document}

\title{A lab scale experiment for keV sterile neutrino search}

\author{Y.C.~Lee$^{a,b}$ \and  H.B.~Kim$^{a,b}$ \and H.L.~Kim$^{a}$ \and S.K.~Kim$^{b}$ \and Y.H.~Kim$^{a,c,d}$ \and D.H.~Kwon$^{a,d}$ \and H.S.~Lim$^{a}$ \and H.S.~Park$^{c}$ \and K.R.~Woo$^{a,d}$ \and Y.S.~Yoon$^{c}$}

\institute{
$^{a}$ Center for Underground Physics, Institute for Basic Science (IBS), Daejeon 34047, Republic of Korea\\
$^{b}$ Department of Physics and Astronomy, Seoul National University, Seoul 08826, Republic of Korea\\
$^{c}$ Korea Research Institute of Standards and Science (KRISS), Daejeon 34113,  Republic of Korea\\
$^{d}$ IBS school, University of Science and Technology (UST), 34113, Daejeon, Korea\\
\email{skkim@snu.ac.kr, yhk@ibs.re.kr}}

\maketitle

\begin{abstract}

We developed a simple small-scale experiment to measure the beta decay spectrum of $^{3}$H. The aim of this research is to investigate the presence of sterile neutrinos in the keV region. Tritium nuclei were embedded in a 1$\times$1$\times$1 cm$^3$ LiF crystal from the $^6$Li(n,$\alpha$)$^3$H reaction. The energy of the beta electrons absorbed in the LiF crystal was measured with a magnetic microcalorimeter at 40\,mK. We report a new method of sample preparation, experiments, and analysis of $^3$H beta measurements.  
The spectrum of a 10-hour measurement agrees well with the expected spectrum of $^3$H beta decay. The analysis results indicate that this method can be used to search for keV-scale sterile neutrinos. 


\end{abstract}

\section{Introduction}

Low-temperature detectors (LTDs) have become one of the key detector technologies in astroparticle physics applications~\cite{booth96,enssbook,yhkim21}.
For neutrino physics in particular, the fundamental properties of neutrinos have been investigated in several outstanding LTD experiments of neutrinoless double beta decay~\cite{amore,cuore,cupid,cupid-mo}. Moreover, the direct detection of the neutrino mass has been intensively carried out with promising sensitivities from an accurate end-point measurement of the $^{163}$Ho electron capture spectrum using low temperature microcalorimeters~\cite{echo,holmes}. On the other hand, the microcalorimeters together with the source manipulation technologies established decay energy spectroscopy that has been adopted in radionuclide analysis and the study of beta decay spectra~\cite{koehler21,ranitzsch20,sjlee10,koehler13}.

Sterile neutrinos are hypothetical right-handed neutrinos that undergo no fundamental interactions other than gravity in the standard model~\cite{giunti07}.
In recent years, the existence of sterile neutrinos was implied by neutrino oscillation experimental results unexpected from the three-flavor model of ordinary neutrinos~\cite{aguilar2001,aa2018}.
On the other hand, if sterile neutrinos exist with their mass scale in the keV range and a very weak interaction strength with ordinary matter, they are viable dark matter candidates~\cite{boyarsky19,gelmini20}.

In the search for sterile neutrinos in the mass region of several hundred keV, the most stringent experimental limit was obtained by a low-temperature experiment based on superconducting tunneling junction devices~\cite{Friedrich21}. However, for the mass range of 1--100\,keV, only conventional detection methods have been applied with the measurement of beta spectra of $^3$H and $^{63}$Ni~\cite{Aadurashitov17,Holzschuh99,adhikari17}. 
Their experimental bounds have not yet been reached at the level of the low-temperature experiment in this mass region.

This present experiment is  a feasibility study for detecting the presence of sterile neutrinos using a sensitive LTD technology. We employed a LiF crystal with a magnetic microcalorimeter (MMC)~\cite{enss00} readout to measure the beta decay events of $^3$H ions that are embedded inside the crystal. 
We report a new method of source preparation, experimental procedure, and analysis to study sterile neutrinos from the $^3$H beta spectrum.

\begin{figure} [t]   %[htbp] %fig1
\begin{center}
\includegraphics[width=0.65\linewidth, keepaspectratio]{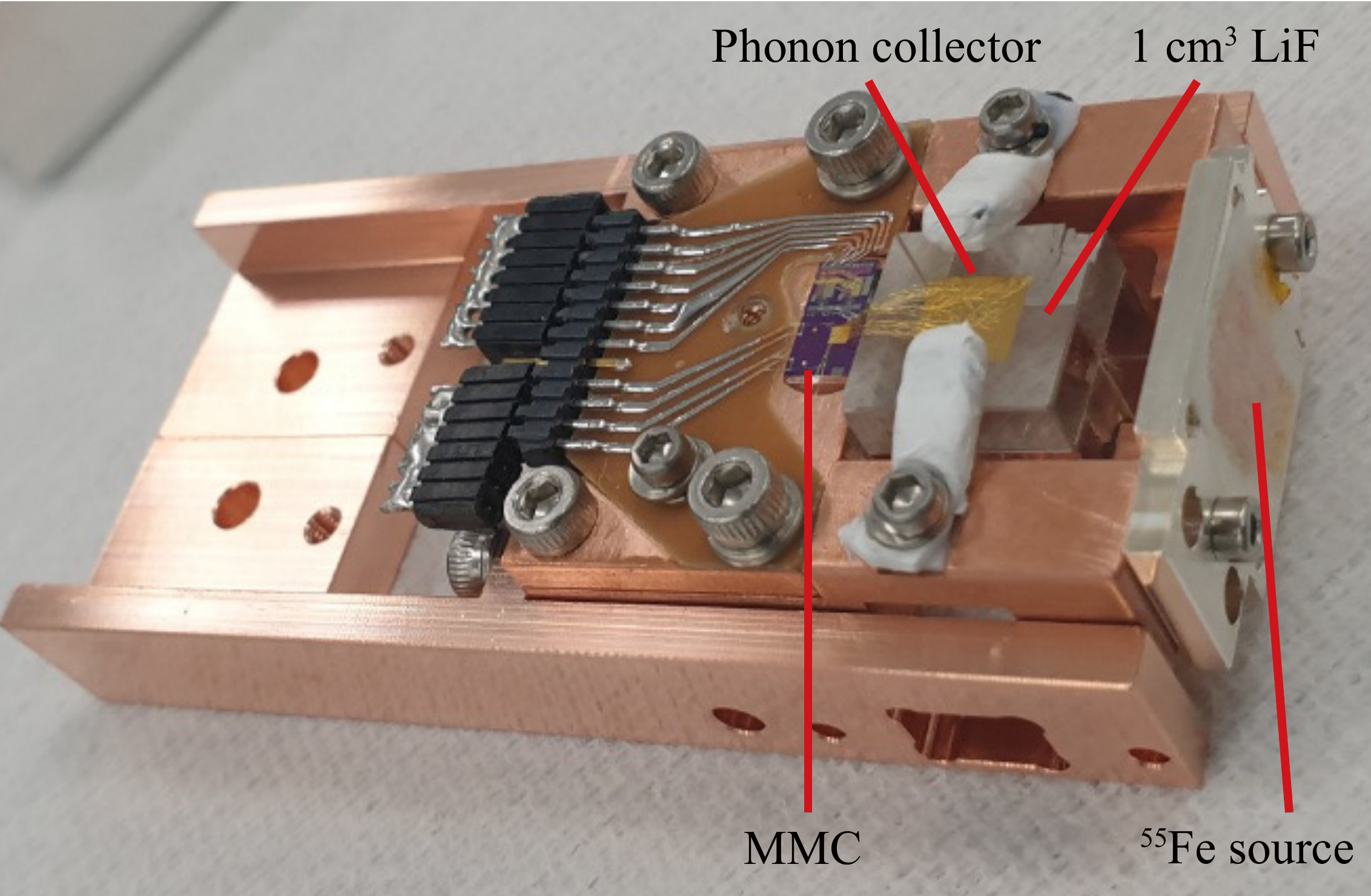}
\caption{Experimental setup with a 1$\times$1$\times$1\,cm$^3$ LiF crystal. 
$^3$H ions are produced in the crystal from the $^6$Li(n,$\alpha$)$^3$H reaction in the bulk.
An MMC placed to the next to the crystal is thermally connected to the crystal via gold bonding wires and a phonon collection film. (Color figure online)}
\end{center}
\label{fig1}
\end{figure}

\section{Experiment}

We developed a small low-temperature thermal calorimeter with a LiF crystal with an MMC readout. A 1$\times$1$\times$1\,cm$^3$ LiF crystal was employed as a hosting material for a $^3$H source and a target absorber for the energetic electrons emitted from $^3$H beta decay.

$^3$H can be embedded in LiF from the $^6$Li(n,$\alpha$)$^3$H reaction. Because the natural abundance of $^6$Li is 7.6\% and  $^6$Li has a large cross-section for thermal neutrons, a considerable amount of $^3$H can be created when a LiF crystal is exposed to a neutron flux. From the reaction, the $^3$H ions are embedded in a LiF crystal and remain in the crystal until they decay to $^3$He with a half-life of 12.32 years via a $\beta$-decay process~\cite{cohen65,luca00}:
\begin{equation}
{^3\mathrm{H}}  \rightarrow {^3\mathrm{He}}  + {\mathrm{e}^-} + {\bar \nu_e}. 
\end{equation}
Electron emission results in a continuous energy spectrum with an end-point at the \emph{Q} value of 18.59\,keV. 

We placed a pure LiF crystal in a neutron flux in laboratory source storage at the Korea Research Institute of Standards and Science (KRISS).  The storage hosts three neutron sources (two AmBe and one $^{252}$Cf) surrounded by polyethylene bricks.
The neutron emissions are expected to be (1--2)$\times$10$^{7}$~n/s from the three sources. 
The storage has a small sample loading volume approximately 17\,cm away from the neutron sources, which is a sufficient distance for the neutrons to be thermalized in the polyethylene bricks.

A 10-day exposure resulted in a rate of 22 $^3$H beta decays per second. Moreover, a Monte Carlo simulation taking neutron thermalization and neutron capture in the LiF crystal into account for $^3$H production showed a similar activity we measured with the present experiment. The mean distance for thermal neutrons to travel in a LiF crystal with natural $^6$Li abundance is 2.3\,mm. This indicates that $^3$H implantation was relatively uniform over the LiF crystal with 1$\times$1$\times$1\,cm$^3$ dimensions. 

The detector composed of  the LiF crystal and an MMC sensor was configured as shown in Fig.~\ref{fig1}. Similar detector setups were developed to survey various molybdate crystals for the AMoRE project~\cite{hlkim2018}.  The crystal positioned in the copper frame was clamped with two Teflon pieces. At the four bottom corners, thin Teflon sheets (not shown in Fig.~\ref{fig1}) were placed between the crystal and the copper frame to minimize heat losses from the crystal to the copper sample holder. 
A 5\,mm$\times$5\,mm$\times$300\,nm gold film was e-beam evaporated on the surface of the crystal after the neutron activation procedure. The gold film is an intermediate thermal component that serves as a phonon collector to make a strong thermal connection to an MMC sensor with gold bonding wires~\cite{hlkim2020,yhkim2004}. The MMC sensor used in the present experiment was fabricated in the Au:Er (1000\,ppm) batch~\cite{sgkim2021}. 
A $^{55}$Fe source was located at the side of the crystal as an internal calibration source. 
The detector setup was attached to an adiabatic demagnetization refrigerator (ADR). The ADR has a 5-cm-thick lead shield around the cryostat to reduce environmental gamma background. We used another $^{241}$Am gamma-ray source located between the cryostat and the lead shield. This source was used only for calibration purposes and was removed when taking the $^3$H beta spectrum.

\begin{figure}[t]%[htbp]   %[htbp] %fig2
\begin{center}
\includegraphics[width=0.7\linewidth, keepaspectratio]{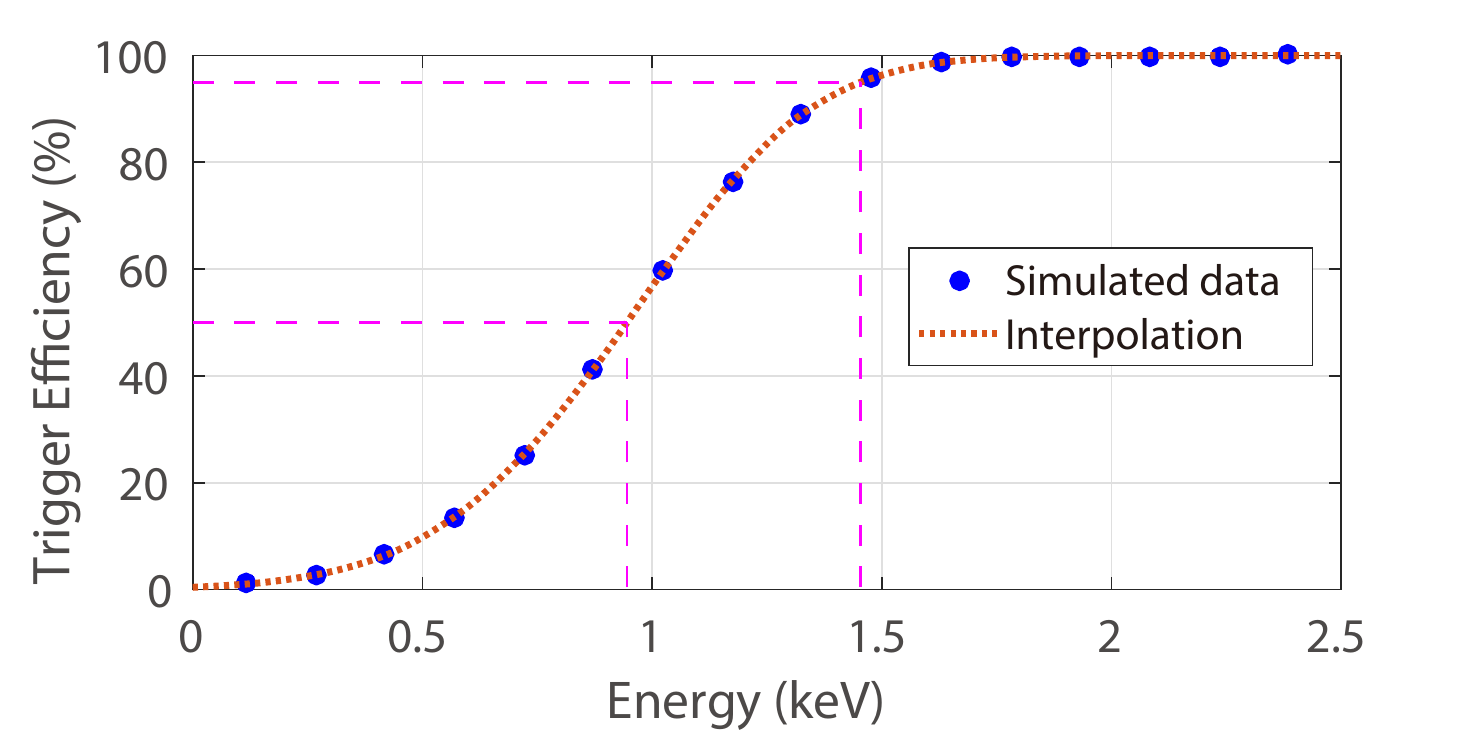}
\caption{Trigger efficiency used for the measurement of the $^3$H beta spectrum with the LiF crystal at 40\,mK. The {\it dashed lines} indicate 50\% and 95\% efficiency levels at 0.95\,keV and and 1.45\,keV, respectively. (Color figure online)}
\label{fig2}
\end{center}
\end{figure}

\begin{figure}[t]%[htbp]   %[htbp] %fig3
\begin{center}
\includegraphics[width=1.0\linewidth, keepaspectratio]{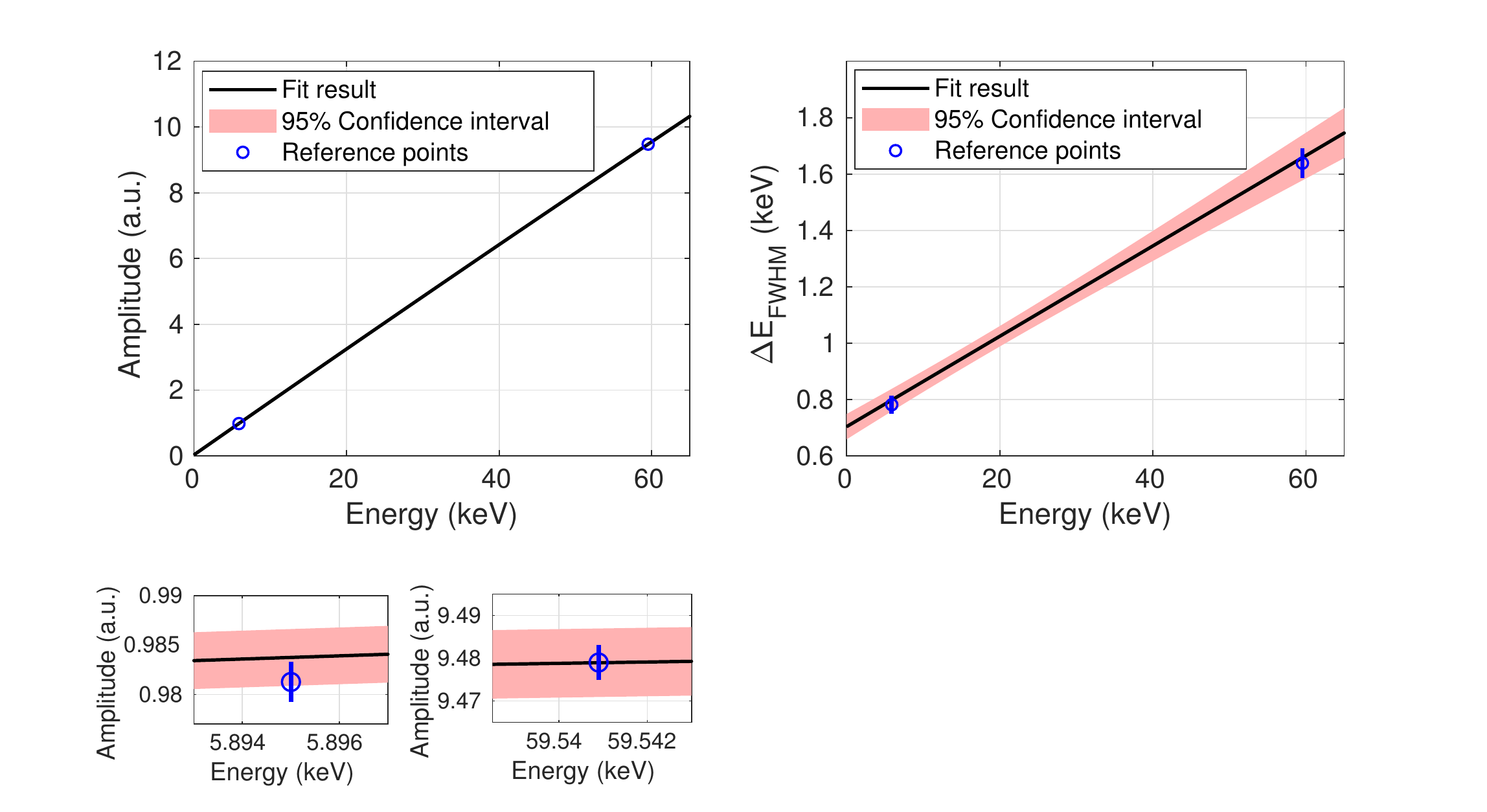}
\caption{The energy calibration ({\it left}) and the energy resolution ({\it right}) found from the total spectrum analysis. 
The reference points indicate the result of individual analysis near the peaks of the calibration sources. (Color figure online)}
\label{fig3}
\end{center}
\end{figure}

\section{Results and Discussions}

The beta spectra were measured three times at 40\,mK. Each measurement lasted 10 hours. The duration was limited by the ADR hold time. The 10-hour data stream was stored in a continuous data acquisition system with a 1 MS/s sampling rate. We developed an offline trigger using a Butterworth filter with a pass band of [30\,kHz 300\,kHz]~\cite{ikim2018}. 
The high-frequency combination was set to the trigger filter to minimize the influence of the low-frequency noise on the trigger signals.  Fig.~\ref{fig2} shows the trigger efficiency with the level set at 4.3 times the standard deviation for the filtered baselines with no signal pulses. We found the trigger efficiency in a pseudo experiment that simulates realistic signal pulses by adding noise signals to a noiseless template scaled to the energy. The template pulse was found from the average signal for the 6\,keV  events from the $^{55}$Fe source and $^{3}$H beta decay. The trigger logic showed  50\% efficiency at 0.95\,keV,  95\% efficiency at 1.45\,keV and nearly 100\% efficiency at 2\,keV.  

The pulse amplitude was determined as a scale factor of the time-domain least-squares method in the comparison of the pulse template and the signals filtered with another combination of pass bands  [310\,Hz 1.09\,kHz]. The pass band frequencies used to determine the pulse amplitude were obtained to maximize the signal-to-noise ratio (the pulse height vs. the rms value of the  baseline).  We also applied a pulse-shape cut using the rms of the difference between the signals and the scaled template. 

The energy  spectrum of  the electron emission from $^3$H beta decay is well understood when no other decay branch is associated, as in the sterile neutrino decay mode. The spectral shape 
$dn / dE_{e}$ 
of the electrons can be expressed as~\cite{kleesiek19}
\begin{equation}
	\frac{dn}{dE_{e}} \propto  F(E_e,Z)p_{e}E_{e}p_{\nu}E_{\nu}\Theta(Q-K_{e}-m_{\nu})	
\label{eq:dnde}
\end{equation}
where $F$ is the relativistic Fermi function,
  $Z=2$ is the atomic number of the daughter, 
$\Theta$ is the unit step function,
$Q$ is the total associated energy of the decay, 
$K_e$ is the kinetic energy of the electron,
and $p_i$, $E_i$, and $m_i$ are the momentum, energy, and mass of the electron and the neutrino respectively. 

In thermal calorimetric detection, the detector measures the total decay energy from the $^3$H beta decay except that taken by the neutrino. The recoil energy of the daughter nucleus has its maximum kinetic energy of 1.72eV~\cite{kleesiek19} and can be considered negligible. The measured energy was set to the kinetic energy of the electrons in the analysis.

\begin{figure} [t]   %[htbp] %fig4
\begin{center}
\includegraphics[width=0.8\linewidth, keepaspectratio]{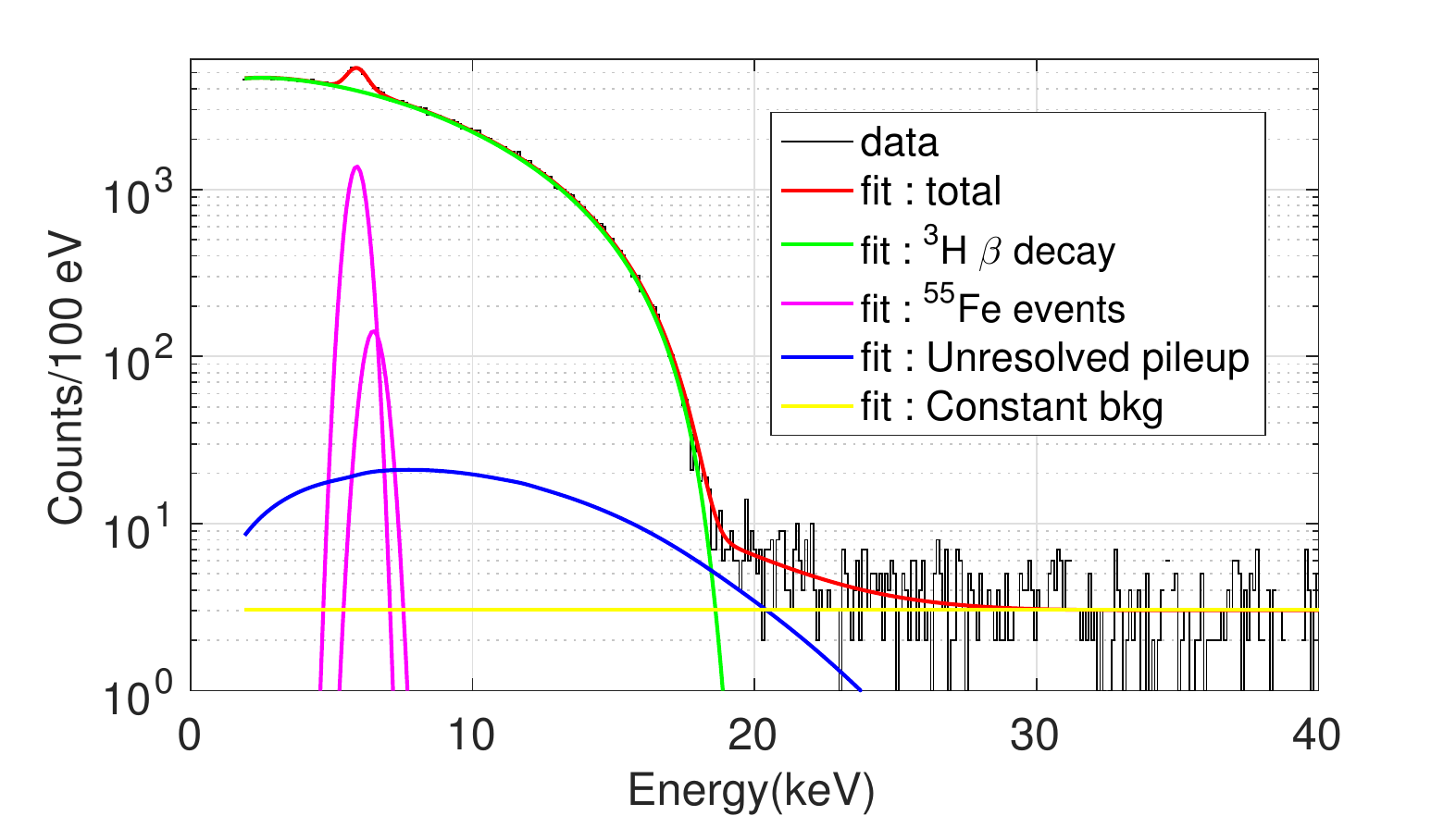}
\caption{The measured and fit spectra. The fit total is the sum of the listed events. (Color figure online)}
\label{fig4}
\end{center}
\end{figure}

Eq.~(\ref{eq:dnde}) was fitted to the measured histogram of the pulse amplitudes in the energy region between 2\,keV and  40\,keV. We fitted the total spectrum by including the events from $^3$H beta decay, $^{55}$Fe source, unresolved pileups, and constant backgrounds in the energy region, as shown in Fig.~\ref{fig4}. The energy calibration and the resolution were considered as fitting parameters to be determined in the spectrum fit. Fig.~\ref{fig3} shows the fit results of energy calibration and resolution.

The fit result of the energy calibration showed a linear behavior, although a general quadratic relation was subject to the fitting parameters. The reference points are the results of the individual analysis near the 6\,keV and 60\,keV regions. They lie in the 95\% confidence interval of the calibration curve. Similarly, the fit result of the energy resolution, assumed to be linear, to the energy agrees well with the individual analysis.  
To add 60\,keV reference data, the fit analysis included the measured spectrum near 60\,keV acquired  with the external $^{241}$Am source. We found a noticeable difference in the low-energy spectra with and without the  $^{241}$Am source due to excess Compton scattering events.

The spectral shape of the unresolved pileups was obtained by considering the probability of pulse shape discrimination for any two events in the spectrum occurring in various time intervals. Approximately \SI{260}{\micro s} timing resolution was found with 99\% rejection probability for the two events with the mean energy of the spectrum.
The total fit result includes a constant level of background events. It influences events with energies larger than 25 keV and has little effect on the $^3$H beta spectrum.  

The resulting spectrum from the fitting analysis matches well with the measured data in the overall energy range between 2\,keV and 40\,keV, as shown in Fig.~\ref{fig4}. The $\chi^2$ goodness of fit, $\chi^2/N$, is equal to 1.047 where $N$ is the degrees of freedom, 423 in our case. Moreover, the p-value of $\chi^2$ is 0.24 for $N$. This suggests that the model fits the observed data at an acceptable level of error.

The spectral shape of the $^3$H decay model used in the fitting analysis was set to follow Eq.~(\ref{eq:dnde}), ignoring the possible decay branch associated with the emission of a massive sterile neutrino. This was mainly because the number of 10-hour events was less than that of the experiment that provides the present experimental bound for sterile neutrinos in the 1-20\,keV region~\cite{Aadurashitov17,Holzschuh99}.

Considering the presence of the beta decay branch with a sterile neutrino, the spectral shape depends on the mixing angle $\theta$ between the massive neutrino and the ordinary three-flavor neutrinos:
\begin{equation}
	\frac{d\lambda}{dE_{e}} = \sin^2\theta\frac{dn}{dE_{e}}\Big|_{m_{\nu_s}}+\cos^2\theta\frac{dn}{dE_{e}}\Big|_{m_{\nu}}
\label{eq:mixing}
\end{equation}
where $m_{\nu_s}$ is the mass of the sterile neutrino. 
Here, the mass of ordinary neutrinos can be set to zero in a keV sterile neutrino search. The mass of the electron neutrino is found to be less than 1\,eV in the recent experiment of a $^3$H end-point measurement~\cite{aker19}. 
The resulting spectral shape of Eq.~(\ref{eq:mixing}) is subject to characteristic distortion at the end point (i.e., $Q - m_{\nu_s}$) of the beta spectrum of the sterile branch.

The presence of sterile neutrinos can be explored with the 10-hour dataset used in this analysis using  Eq.~(\ref{eq:mixing}). A limit bound is expected for the mixing parameters of $\sin^2\theta$ and  $m_{\nu_s}$ for the null result. We plan to expand this method for the other two sets of data. With all the datasets, the analysis is expected to reach a level close to the present up-to-date limit in the 1--16 keV region. 

\section{Conclusion}

The energy spectrum of the electron emission in $^3$H beta decay can be accurately measured with a LiF detector with an MMC readout. Because the detector is sensitive to almost the full energy range of the decay, the detection method allows for the search of sterile neutrinos in a similar mass range. Moreover, it was a successful demonstration of $^3$H implantation in a LiF crystal, having a reasonable decay rate of $^3$H ions. A larger number of crystals can be prepared with the control of the activity. 

As a next step of the experiment, we plan to increase the channel number and the event rate in a crystal. Furthermore, future experiments should be carried out in a dilution refrigerator. Data acquisition does not have to be interrupted as it was in the ADR from the magnet cycles but can be settled for long-term measurement. Measurements performed at a lower temperature increase the signal amplitudes. This will increase the sensitivity to reject the pileup signals as well as the energy resolution. As a rule of thumb, a measurement for 12 channel$\cdot$month with 40\,Bq/channel reaches the mixing amplitude $\sin^2\theta$ of $10^{-4}$--$10^{-3}$ in the keV mass region.

\begin{acknowledgements}
This research is supported by Grant No. IBS-R016-A2 and NRF-2021R1A2C3010989.
\end{acknowledgements}


\begin{thebibliography}{99}

\bibitem{booth96}
N. E. Booth, B. Cabrera, E. Fiorini, {\it Annu. Rev. Nucl. Part. Sci.} \textbf{46} 471 (1996).

\bibitem{enssbook}
C. Enss (Ed.), {\it Cryogentic Particle Detection}  (Springer, Beriln, 2005).

\bibitem{yhkim21}
Y.H. Kim, S.J. Lee, B.S. Yang,  ``Superconducting detectors for rare event search experiments'',  submitted to {\it Supercond. Sci. Technol.} (2021).

\bibitem{amore}
V. Alenkov, H.W. Bae, J. Beyer et al., {\it Eur. Phys. J. C} \textbf{79} 791 (2019).

\bibitem{cuore}
D.Q. Adams et al., {\it Phys. Rev. Lett.} \textbf{124} 122501 (2020).

\bibitem{cupid}
D.R.Artusa et al., {\it Phys. Lett. B} \textbf{767}, 321, (2017).

\bibitem{cupid-mo}
E. Armengaud, C. Augier, A.S. Barabash et al., {\it Eur. Phys. J. C} \textbf{80} 44 (2020).

\bibitem{echo}
L Gastaldo et al., {\it J. Low Temp. Phys.} \textbf{176} 876--884 (2014).

\bibitem{holmes}
B Alpert et al., {\it Eur. Phys. J. C} \textbf{75} 112 (2015).

\bibitem{koehler21}
K.E. Koehler, {\it Appl. Sci.} \textbf{11(9)} 4044 (2021).

\bibitem{ranitzsch20} 
P.C.-O. Ranitzsch et al., {\it J. Low Temp. Phys.} \textbf{199} 441--450 (2020).

\bibitem{sjlee10}
S.J. Lee et al., {\it J. Phys. G: Nucl. Part. Phys.} \textbf{37} 055103 (2010).

\bibitem{koehler13}
K.E. Koehler et al., {\it IEEE. Trans. Nucl. Sci.} \textbf{60}, 2 (2013).

\bibitem{giunti07}
C. Giunti and C.W. Kim, {\it Fundamentals of Neutrino Physics and Astrophysics}, ill. edn. (Oxford University Press, Oxford, 2007), \textbf{p. 180}.

\bibitem{aguilar2001}
A Aguilar et al., {\it Phys. Rev. D} \textbf{64} 112007 (2001)

\bibitem{aa2018}
A A Aguilar-Arevalo et al.,  {\it Phys. Rev. Lett.} \textbf{121} 221801 (2018)

\bibitem{boyarsky19}
A. Boyarsky, M. Drewes, T. Lasserre, S. Mertens, O. Ruchayskiy, {\it Prog. Part. Nucl. Phys.} \textbf{104}, 1, (2019).

\bibitem{gelmini20}
G.B. Gelmini, P. Lu, V. Takhistov, {\it Phys. Lett. B} \textbf{800}, 10, (2020).

\bibitem{Friedrich21}
S. Friedrich et al., {\it Phys. Rev. Lett.} \textbf{126}, 021803, (2021). 

\bibitem{Aadurashitov17}
J.N. Abdurashitov, A.I. Belesev, V.G. Chernov et al., {\it JETP Lett.} \textbf{105}, 753, (2017).

\bibitem{Holzschuh99}
E. Holzschuh, W. Kundig, L. Palermo, H. Stussi, P. Wenk, {\it Phys. Lett. B} \textbf{451}, 247, (1999).

\bibitem{adhikari17}
R. Adhikari et al., {\it JCAP} \textbf{1701}, 025, (2017).

\bibitem{enss00} 
C. Enss et al., {\it J. Low Temp. Phys.} \textbf{121} 137--176 (2000).

\bibitem{cohen65}
H. Cohen, W.S. Diethorn,, {\it Phys. Stat. Sol.} \textbf{9}, 251, (1965).

\bibitem{luca00}
L.L. Luca, M.P. Unterweger, {\it J. Res. Natl. Inst. Stand. Technol.} \textbf{105}, 541, (2000).

\bibitem{hlkim2018}
H.L. Kim et al., {\it IEEE. Trans. Nucl. Sci.} \textbf{65}, 2 (2018).

\bibitem{hlkim2020}
H.L. Kim et al., {\it Nucl. Inst. Method A} \textbf{954}, 21 (2020).

\bibitem{yhkim2004}
Y.H. Kim et al., {\it Nucl. Inst. Method A} \textbf{520}, 208 (2004).

\bibitem{sgkim2021}
S.G. Kim, et al., {\it IEEE T. Appl. Supercon.}  \textbf{31}, 1--5 (2021).

\bibitem{ikim2018}
I. Kim, et al., {\it J. Low Temp. Phys.} \textbf{193} 1190--1198 (2018).

\bibitem{kleesiek19}
M. Kleesiek, J. Behrens, G. Drexlin et al.{\it Eur. Phys. J. C} \textbf{79}, 204, (2019).

\bibitem{aker19}
M. Aker et al.,  {\it Phys. Rev. Lett.} \textbf{123} 221802 (2019).



\end{thebibliography}
\end{document}